\documentclass[prd,amsfont,showkeys,notitlepage]{revtex4-1}

\usepackage[]{graphicx}
\usepackage{amsmath,amssymb}

\usepackage[breaklinks=true]{hyperref}

\newcommand{\E}{E_{0}}
\newcommand{\rhos}{\rho_{s,600}}
\newcommand{\meanLnA}{\left<\log{A}\right>}
\newcommand{\Rm}{r_{\text{M}}}
\newcommand{\exper}{{\text{exp.}}}
\newcommand{\mcP}{{\text{p}}}
\newcommand{\mcFe}{{\text{Fe}}}

\newcommand{\degr}{^{\circ}}
\newcommand{\depthunit}{g/cm$^2$}
\newcommand{\xmax}{x_{\text{max}}}

\newcommand{\corsika}{\textsc{corsika}}
\newcommand{\QGS}{QGSJet01d}
\newcommand{\QGSII}{QGSJet-II-04}
\newcommand{\SIBYLL}{SIBYLL-2.1}
\newcommand{\EPOS}{EPOS-LHC}

\begin{document}

\title{Mass composition of cosmic rays with energy above $10^{17}$~eV according to surface detectors of the Yakutsk EAS array}

\author{A. V. Glushkov}

\author{A. Sabourov}
\email{tema@ikfia.sbras.ru}

\affiliation{Yu. G. Shafer Institute of cosmophysical research and aeronomy}
\address{677980, Lenin Ave. 31, Yakutsk, Russia}

\begin{abstract}
    We discuss the lateral distribution of charged particles in extensive air showers with energy above $10^{17}$~eV measured by surface scintillation detectors of Yakutsk EAS array. The analysis covers the data obtained during the period from 1977 to 2013. Experimental values are compared to theoretical predictions obtained with the use of {\corsika} code within frameworks of different hadron interaction models. The best agreement between theory and experiment is observed for \QGS{} and \QGSII{} models. A change in the cosmic ray mass composition towards proton is observed in the energy range $(1 - 20) \times 10^{17}$~eV.
\end{abstract}



\maketitle

\section{Introduction}

The mass composition of cosmic rays (CR) with energy $\E \ge 10^{17}$~eV is still not known precisely despite the fact that it has been actively studied on world extensive air shower (EAS) arrays for more than 40~years.~\cite{Greider(2010)}. This research is based on various EAS parameters that are sensitive to the CR mass composition. For measurement of these parameters the Yakutsk experiment utilizes the lateral distribution functions (LDF) of electron, muon and Cherenkov components of EAS (see e.g.~\cite{Glushkov:PhD(1982), Yakutsk:Bull.Acad.Sci.USSR(1986), Yakutsk:Phys.Atom.Nucl.(2000), Yakutsk:JETPLett.(2013), Yakutsk:Astropart.Phys.(2012)}). One of the key chracteristics that could be estimated on a ground array is the depth of maximum of a shower cascade curve ($\xmax$) which is connected with the atomic number $A$ of primary CR particle with the relation:

\begin{equation}
    \log{A} = \log{56} \cdot \frac{\xmax^\mcP - \xmax^\exper}{\xmax^\mcP - \xmax^\mcFe}\text{,}
    \label{eq:LnA}
\end{equation}
where $\xmax^\exper$ is experimentally measured value and $\xmax^\mcP, \xmax^\mcFe$ are values obtained via calculations performed for primary protons and iron nuclei. Here, one cannot do without theoretical notion of EAS development. Earlier, the lateral distribution of signal in surface scintillation detectors of the Yakutsk array have been calculated~\cite{Yakutsk:JETPLett.(2014)}. The calculations were performed with the use of {\corsika} code~\cite{CORSIKA} for primary particles with energies $\E \ge 10^{17}$~eV within the framework of \QGS~\cite{QGSJet01}, \QGSII~\cite{QGSJetII}, \SIBYLL~\cite{SIBYLL} and \EPOS~\cite{EPOS} models. FLUKA package~\cite{FLUKA:Proc.AIP(2007)} was chosen for treatment of low-energy interactions. Further, we compare LDFs predicted by these models with experimental data obtained during the period of continuous observation lasted from 1977 to 2013.

\section{Results and discussion}

\begin{figure}[htb]
    \centering
    \includegraphics[width=0.65\textwidth, clip]{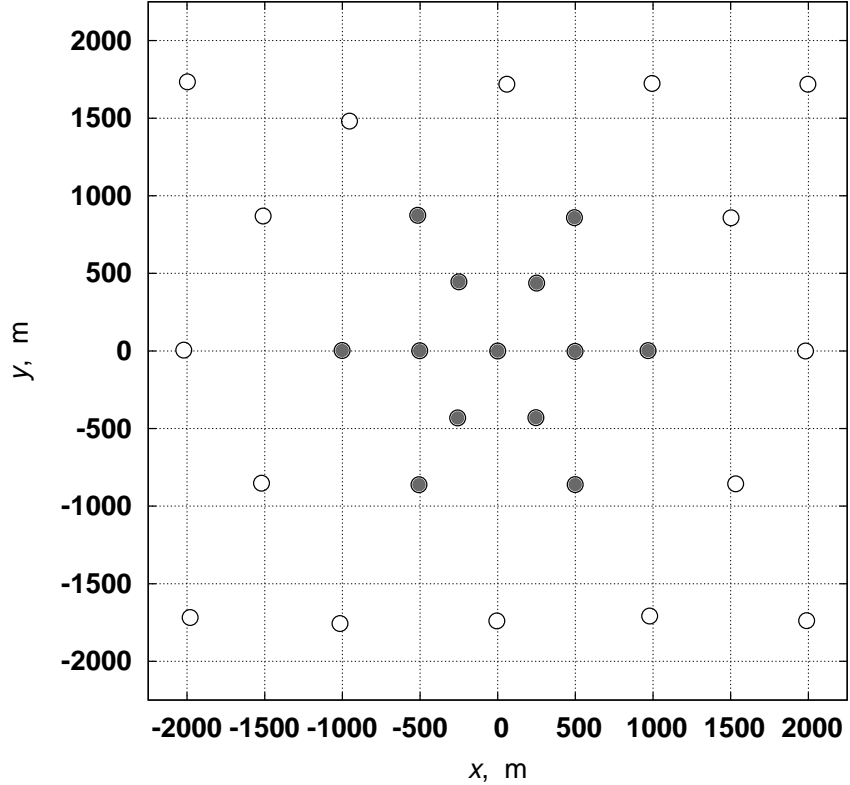}
    \caption{The layout of master stations of Yakutsk EAS array. Stations whose indications were used in the analysis are shaded with gray color.}
    \label{fig:array}
\end{figure}

\begin{figure}[htb]
    \centering
    \includegraphics[width=0.65\textwidth]{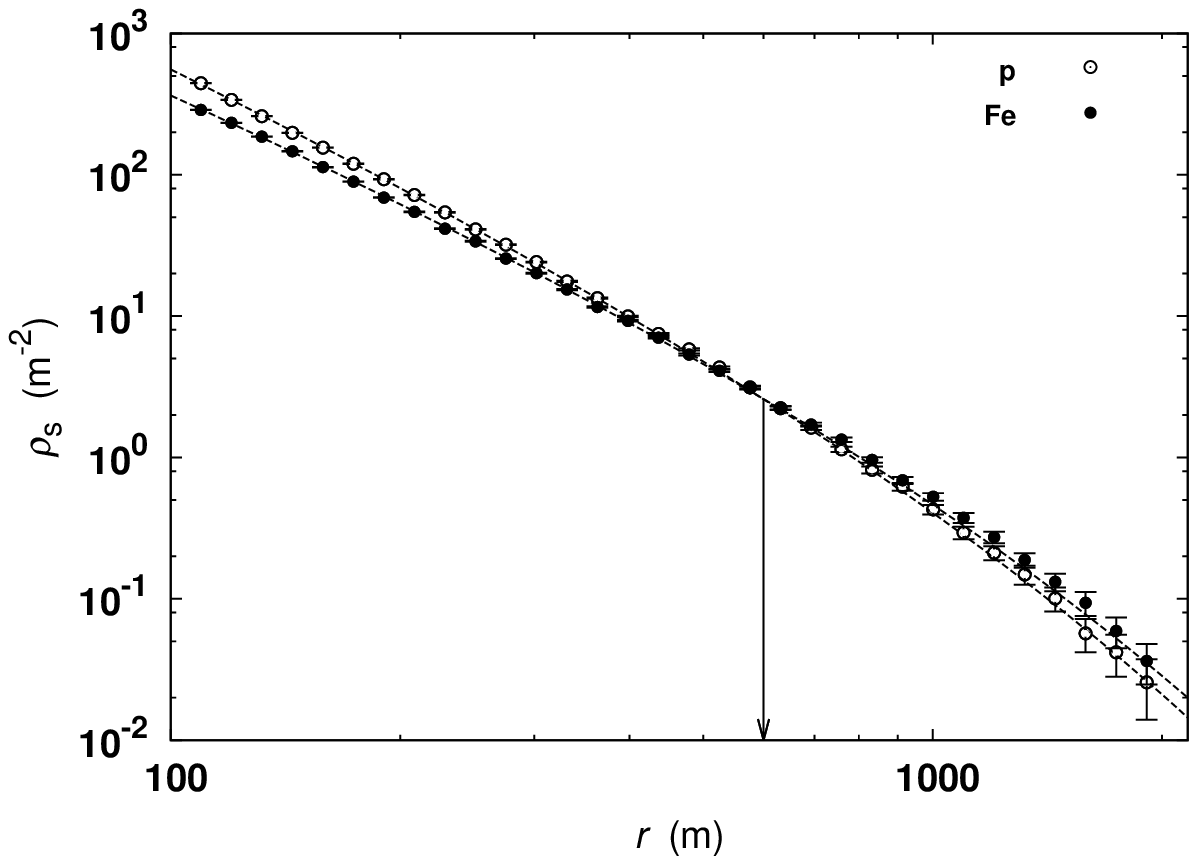}
    \caption{LDFs of charged particles in showers with $\E = 10^{18}$~eV and $\cos{\theta} = 0.9$ for primary protons and iron nuclei according to \QGSII{} model. Dashed lines represent approximations, arrow indicates the value of density at $600$~m from shower axis.}
    \label{fig:1}
\end{figure}

EAS events covered by the analysis have zenith angles of arrival direction $\theta \le 25.8\degr$ ($\left<\cos{\theta}\right> = 0.95$). Only $13$ stations located around the center of the array were used to construct LDF. With the central station they form 6 master triangles with $500$~m side (the ``small master''~--- SM) and 6~--- with $1000$~m side (``large master''~--- LM), as shown of Fig.\ref{fig:array}. These stations have two scintillation detectors ($2 \times 2$~m$^2$) operating in coincidence mode. According to~\cite{Yakutsk:JETPLett.(2014)}, the energy of primary particles was determined by relations:
\begin{eqnarray}
    \E = & (3.40 \pm 0.18) \times 10^{17} \cdot (\rhos(0\degr))^{1.017}\text{,\quad eV}
    \label{eq:Energy}\\
    &  \rhos(0\degr) = \rhos(\theta) \cdot \exp{\frac{(\sec{\theta} - 1) \cdot 1020}{\lambda_{\rho}}}\text{,\quad m$^{-2}$}\\
    \label{eq:rho_vert}
    & \lambda_{\rho} = 415 \pm 5 \text{,\quad\depthunit}
    \label{eq:FreePath}
\end{eqnarray}
where $\rhos(\theta)$ is the density of shower particles measured by surface scintillation detectors at the distance $r = 600$~m from shower axis. The relation~(\ref{eq:Energy}) unambiguously connects $\rhos(0\degr)$ with $\E$ at any CR composition, since at $\sim 600$ the LDFs of charged particles intercross each other. It is demonstrated on Fig.\ref{fig:1} where two simulation results are shown obtained for protons and iron nuclei with $\E = 10^{18}$~eV and $\cos{\theta} = 0.9$.

Geometrical reconstruction of the considered showers was performed with the use of function
\begin{equation}
    f_s(r, \theta) = \rhos(\theta) \cdot \left(\frac{600 + r_1}{r + r_1}\right)^{a} \cdot \left(\frac{600 +\Rm}{r + \Rm}\right)^{b - a}\text{,}
    \label{eq:LDF_SD}
\end{equation}
where $a = 1$, $r_1 = 0$, $\Rm$~--- is the Moliere radius which depends on the air temperature $t$ ($\degr$C) and air pressure $P$ (mbar):
\begin{equation}
    \Rm = \simeq \frac{7.5 \times 10^4}{P} \cdot \frac{t + 273}{273}~\text{,\quad m.}
    \label{eq:Moliere}
\end{equation}
The $\Rm$ value is determined for every registered shower (for Yakutsk $\left<t\right> \simeq -18\degr$C and $\left<\Rm\right> \simeq 70$~m). In equation~(\ref{eq:LDF_SD}) $b$ is the parameter defined earlier~\cite{Yakutsk:Ya.Fil.SB.Acad.Sci.USSR(1976)}:
\begin{equation}
    b = 1.38 + 2.16 \times \cos{\theta} + 0.15 \times \log_{10}{\rhos(\theta})\text{.}
    \label{eq:LDF_b}
\end{equation}

The final analysis includes showers whose errors of axis reconstruction do not exceed $20-30$~m for SM and $50$~m~--- for LM. Mean LDFs were constructed in energy bins with logarithmic step $h = \Delta \log_{10}{\E} = 0.2$ with a subsequent shift by $0.5h$, in order to examine in detail the agreement between the experiment and a given model. During construction of mean LDFs, particle densities were multiplied by normalization ratio $\left<\E\right> / \E$ ($\left<\E\right>$ being the mean energy in a group) and averaged within radial bins $\Delta\log_{10}{r} = 0.04$. Mean particle densities were determined with the formula:
\begin{equation}
    \left<\rho_s(r_i)\right> = \frac{\sum_{k = 1}^{N} \rho_k(r_i)}{N}\text{,}
    \label{eq:MeanDensity}
\end{equation}
where $N$ is the number of readings from detectors within axis distance ranges $(\log_{10}{r_i}, \log_{10}{r_i} + 0.04)$.

The resulting LDFs were approximated with the function
\begin{equation}
    \rhos(r,\theta) = f_s(r,\theta) \cdot \left(\frac{600 + r_2}{r + r_2}\right)^{12}\text{,}
    \label{eq:LateralDistribution}
\end{equation}
where $a = 2$, $\Rm = 10$, $r_1 = 8$ and $r_2 = 10^{4}$~m. Here, the $\Rm$ has become a formal parameter. In the aggregate with other parameters from (\ref{eq:LateralDistribution}) it provides the best agreement with densities (\ref{eq:MeanDensity}) in the whole range $20-2000$~m from the axis. The best fit values for $\rhos(\theta)$ and $b$ in individual groups were derived through $\chi^2$ minimization.

\begin{figure}[htb]
    \centering
    \includegraphics[width=0.65\textwidth]{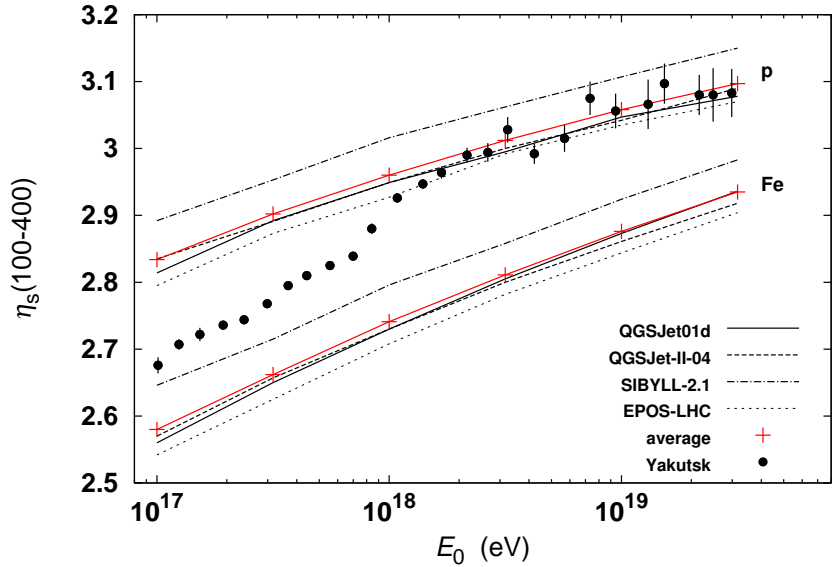}
    \caption{Local steepness of the surface detector response LDF in the distance range $(100-400)$~m  in showers with $\left<\cos{\theta}\right> = 0.95$.}
    \label{fig:eta}
\end{figure}

The parameter $b$ reflects the steepness of LDF, which is sensitive to CR mass composition. On Fig.\ref{fig:eta} the local steepness of LDF is shown
\begin{equation}
    \eta_s(100-400) = \frac{\log_{10}{\rho_s(100)} - \log_{10}{\rho_s(400)}}{\log_{10}(400/100)}
\end{equation}
in the ditance range $100-400$~m. Its value is close to $b$ but it can be measured for all energies. During the parametrization of (\ref{eq:LateralDistribution}), particle densities outside the specified range were omitted. Lines represent expected values predicted by four models used in {\corsika} simulations. 200 showers were simulated for each set of primary parameters (mass of primary particle, energy, zenith angle). In order to speed-up the simulation the thin-sampling mechanism was activated in all versions of {\corsika} code with the following parameters:  $\epsilon_i / E_0 \in [3.16 \cdot 10^{-6}, 10^{-5}]$ and $w_{\text{max}} \in [10^4, 3.16 \cdot 10^{6}]$, depending on the primary energy. During calculation of particle density we considered the response of scintillation detectors from muons, gamma-photons and electrons~\cite{Yakutsk:JETPLett.(2014)}.

On Fig.\ref{fig:eta} the dependency obtained by averaging of predictions of all models is shown with crosses. It is closest to \QGS{} and \QGSII{} models and provides the possibility to estimate the mass composition of primary particles from the relation:

\begin{equation}
    \meanLnA = w_{\mcP} + w_{\mcFe} \cdot \log{56}\text{.}
    \label{eq:meanLnA}
\end{equation}
Here $w_{\mcP} = w_{\mcFe}$ and $w_{\mcFe} = \meanLnA / \log{56}$. From this notion we have:
\begin{equation}
    w_{\mcFe} = \frac{d_{\exper} - d_{\mcP}}{d_{\mcFe} - d_{\mcP}}\text{,}
    \label{eq:wFe}
\end{equation}
where $d = \eta_s(100-400)$, the values obtained in experiment and in simulation.

\begin{figure}[htb]
    \centering
    \includegraphics[width=0.65\textwidth]{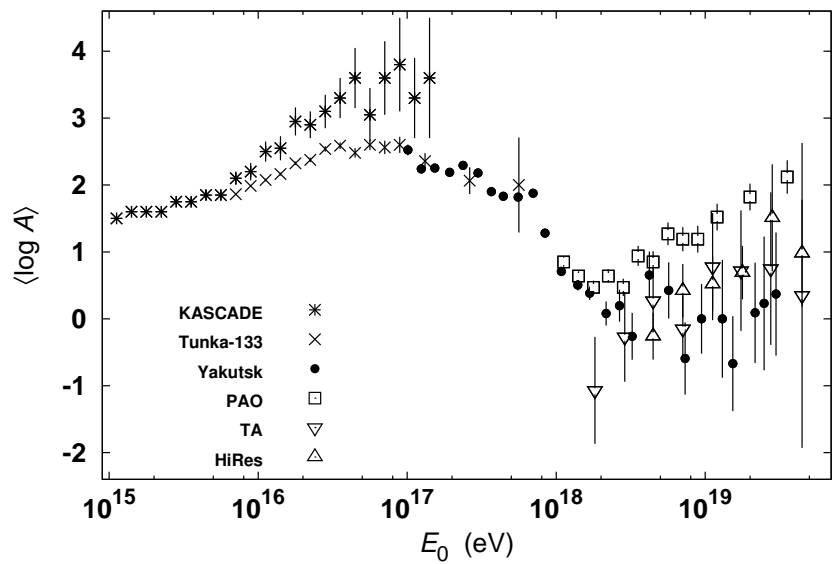}
    \caption{Energy dependence of the CR mass composition according to several EAS experiments.}
    \label{fig:meanLnA}
\end{figure}

On Fig.\ref{fig:meanLnA} energy dependencies are shown obtained by various EAS experiments. Dark circles represent our estimation based on (\ref{eq:meanLnA}) and (\ref{eq:wFe}) for averaged dependency shown on Fig.\ref{fig:eta}. Stars represent the data of KASCADE during the period from May 1998 to December 1999~\cite{KASCADE:Proc27thICRC(2001)}. With crosses are denoted the data of Tunka-133 obtained from Cherenkov light LDF~\cite{Tunka:Astropart.Phys.(2013)}. White squares~--- PAO~\cite{AUGER(2010)}, upward triangles~--- HiRes~\cite{HiRes:Phys.Rev.Lett.(2010)}, downward triangles~--- TA~\cite{TA(2011)}. The last three sets were obtained from the $\xmax$ values presented in~\cite{AUGER(2010),HiRes:Phys.Rev.Lett.(2010),TA(2011)} according to averaged values $\left<\xmax(\E)\right>$ for abovementioned hadron interaction models. All results are roughly consistent with each other except for $\meanLnA$ at $\E \ge 2 \times 10^{18}$~eV resulting from the PAO data~\cite{AUGER(2010)}.

\section{conclusion}

Long-term observation of ultra-high energy cosmic rays at the Yakutsk EAS array and comparison of experimental results to simulation~\cite{Yakutsk:JETPLett.(2014)} have provided the opportunity to estimate the CR mass composition in the energy range $\E \simeq 10^{17}-10^{18}$~eV where experimental data are notably sparse. On Fig.\ref{fig:meanLnA} a rapid change of the mass composition with energy is seen in energy range $(1-20) \times 10^{17}$~eV towards lighter nuclei. This is probably due to a transition from galactic CR to extragalactic. One may assume that at $\E \ge 2 \times 10^{18}$~eV primary particles are mainly protons. However it is a bit premature to make such a strict conclusion. Further research is required to give a definitive answer.

\acknowledgements

Simulations were performed on the \textsl{Arian Kuz'min} supercomputer of the North-Eastern Federal University (Yakutsk).

The work is financially supported by Russian Academy of Science within the program ``Fundamental properties of matter and astrophysics'', by RFBR grant 13-02-12036~ofi-m-2013 and by grant of President of the Republic Sakha (Yakutia) to young scientists, specialists and students to support research. 

\bibliographystyle{aipnum4-1}
\bibliography{masscomp2014}

\begin{thebibliography}{19}%
\makeatletter
\providecommand \@ifxundefined [1]{%
 \@ifx{#1\undefined}
}%
\providecommand \@ifnum [1]{%
 \ifnum #1\expandafter \@firstoftwo
 \else \expandafter \@secondoftwo
 \fi
}%
\providecommand \@ifx [1]{%
 \ifx #1\expandafter \@firstoftwo
 \else \expandafter \@secondoftwo
 \fi
}%
\providecommand \natexlab [1]{#1}%
\providecommand \enquote  [1]{``#1''}%
\providecommand \bibnamefont  [1]{#1}%
\providecommand \bibfnamefont [1]{#1}%
\providecommand \citenamefont [1]{#1}%
\providecommand \href@noop [0]{\@secondoftwo}%
\providecommand \href [0]{\begingroup \@sanitize@url \@href}%
\providecommand \@href[1]{\@@startlink{#1}\@@href}%
\providecommand \@@href[1]{\endgroup#1\@@endlink}%
\providecommand \@sanitize@url [0]{\catcode `\\12\catcode `\$12\catcode
  `\&12\catcode `\#12\catcode `\^12\catcode `\_12\catcode `\%12\relax}%
\providecommand \@@startlink[1]{}%
\providecommand \@@endlink[0]{}%
\providecommand \url  [0]{\begingroup\@sanitize@url \@url }%
\providecommand \@url [1]{\endgroup\@href {#1}{\urlprefix }}%
\providecommand \urlprefix  [0]{URL }%
\providecommand \Eprint [0]{\href }%
\providecommand \doibase [0]{http://dx.doi.org/}%
\providecommand \selectlanguage [0]{\@gobble}%
\providecommand \bibinfo  [0]{\@secondoftwo}%
\providecommand \bibfield  [0]{\@secondoftwo}%
\providecommand \translation [1]{[#1]}%
\providecommand \BibitemOpen [0]{}%
\providecommand \bibitemStop [0]{}%
\providecommand \bibitemNoStop [0]{.\EOS\space}%
\providecommand \EOS [0]{\spacefactor3000\relax}%
\providecommand \BibitemShut  [1]{\csname bibitem#1\endcsname}%
\let\auto@bib@innerbib\@empty
\bibitem [{\citenamefont {Greider}(2010)}]{Greider(2010)}%
  \BibitemOpen
  \bibfield  {author} {\bibinfo {author} {\bibfnamefont {P.~K.~F.}\
  \bibnamefont {Greider}},\ }\href {\doibase 10.1007/978-3-540-76941-5} {\emph
  {\bibinfo {title} {{Extensive Air Showers: High Energy Phenomena and
  Astrophysical Aspects~--- A Tutorial, Reference Manual and Data Book}}}}\
  (\bibinfo  {publisher} {{Springer Berlin Heidelberg}},\ \bibinfo {year}
  {2010})\ p.\ \bibinfo {pages} {1115}\BibitemShut {NoStop}%
\bibitem [{\citenamefont {Glushkov}(1982)}]{Glushkov:PhD(1982)}%
  \BibitemOpen
  \bibfield  {author} {\bibinfo {author} {\bibfnamefont {A.~V.}\ \bibnamefont
  {Glushkov}},\ }\emph {\bibinfo {title} {{Lateral distribution and total flux
  of Cherenkov radiation from EAS with primary energy $\E \ge 10^{17}$~eV}}},\
  \href@noop {} {Ph.D. thesis},\ \bibinfo  {school} {{SINP MSU}} (\bibinfo
  {year} {1982}),\ \bibinfo {note} {{(in Russian)}}\BibitemShut {NoStop}%
\bibitem [{\citenamefont {Glushkov}\ \emph {et~al.}(1986)\citenamefont
  {Glushkov}, \citenamefont {Dedenko}, \citenamefont {Efimov}, \citenamefont
  {Efremov}, \citenamefont {T.} \emph
  {et~al.}}]{Yakutsk:Bull.Acad.Sci.USSR(1986)}%
  \BibitemOpen
  \bibfield  {author} {\bibinfo {author} {\bibfnamefont {A.~V.}\ \bibnamefont
  {Glushkov}}, \bibinfo {author} {\bibfnamefont {L.~G.}\ \bibnamefont
  {Dedenko}}, \bibinfo {author} {\bibfnamefont {N.~N.}\ \bibnamefont {Efimov}},
  \bibinfo {author} {\bibfnamefont {N.~N.}\ \bibnamefont {Efremov}}, \bibinfo
  {author} {\bibfnamefont {M.~I.}\ \bibnamefont {T.}},  \emph {et~al.},\
  }\href@noop {} {\bibfield  {journal} {\bibinfo  {journal} {{Bull. Acad. Sci.
  USSR}}\ }\bibinfo {series} {{Phys. Ser.}},\ \textbf {\bibinfo {volume}
  {55}},\ \bibinfo {pages} {2166} (\bibinfo {year} {1986})},\ \bibinfo {note}
  {{(in Russian)}}\BibitemShut {NoStop}%
\bibitem [{\citenamefont {Glushkov}\ \emph {et~al.}(2000)\citenamefont
  {Glushkov}, \citenamefont {Pravdin}, \citenamefont {Sleptsov}, \citenamefont
  {Sleptsova},\ and\ \citenamefont {Kalmykov}}]{Yakutsk:Phys.Atom.Nucl.(2000)}%
  \BibitemOpen
  \bibfield  {author} {\bibinfo {author} {\bibfnamefont {A.}~\bibnamefont
  {Glushkov}}, \bibinfo {author} {\bibfnamefont {M.}~\bibnamefont {Pravdin}},
  \bibinfo {author} {\bibfnamefont {I.}~\bibnamefont {Sleptsov}}, \bibinfo
  {author} {\bibfnamefont {V.}~\bibnamefont {Sleptsova}}, \ and\ \bibinfo
  {author} {\bibfnamefont {N.}~\bibnamefont {Kalmykov}},\ }\href {\doibase
  10.1134/1.1307469} {\bibfield  {journal} {\bibinfo  {journal} {Phys. of Atom.
  Nucl.}\ }\textbf {\bibinfo {volume} {63}},\ \bibinfo {pages} {1477} (\bibinfo
  {year} {2000})}\BibitemShut {NoStop}%
\bibitem [{\citenamefont {Glushkov}\ and\ \citenamefont
  {Saburov}(2014)}]{Yakutsk:JETPLett.(2013)}%
  \BibitemOpen
  \bibfield  {author} {\bibinfo {author} {\bibfnamefont {A.~V.}\ \bibnamefont
  {Glushkov}}\ and\ \bibinfo {author} {\bibfnamefont {A.~V.}\ \bibnamefont
  {Saburov}},\ }\href {\doibase 10.1134/S0021364013230057} {\bibfield
  {journal} {\bibinfo  {journal} {{JETP Lett.}}\ }\textbf {\bibinfo {volume}
  {98}},\ \bibinfo {pages} {589} (\bibinfo {year} {2014})}\BibitemShut
  {NoStop}%
\bibitem [{\citenamefont {Berezhko}, \citenamefont {Knurenko},\ and\
  \citenamefont {Ksenofontov}(2012)}]{Yakutsk:Astropart.Phys.(2012)}%
  \BibitemOpen
  \bibfield  {author} {\bibinfo {author} {\bibfnamefont {E.~G.}\ \bibnamefont
  {Berezhko}}, \bibinfo {author} {\bibfnamefont {S.~P.}\ \bibnamefont
  {Knurenko}}, \ and\ \bibinfo {author} {\bibfnamefont {L.~T.}\ \bibnamefont
  {Ksenofontov}},\ }\href {\doibase 10.1016/j.astropartphys.2012.04.014}
  {\bibfield  {journal} {\bibinfo  {journal} {{Astropart. Phys.}}\ }\textbf
  {\bibinfo {volume} {36}},\ \bibinfo {pages} {31} (\bibinfo {year}
  {2012})}\BibitemShut {NoStop}%
\bibitem [{\citenamefont {Glushkov}, \citenamefont {Pravdin},\ and\
  \citenamefont {Saburov}(2014)}]{Yakutsk:JETPLett.(2014)}%
  \BibitemOpen
  \bibfield  {author} {\bibinfo {author} {\bibfnamefont {A.~V.}\ \bibnamefont
  {Glushkov}}, \bibinfo {author} {\bibfnamefont {M.~I.}\ \bibnamefont
  {Pravdin}}, \ and\ \bibinfo {author} {\bibfnamefont {A.}~\bibnamefont
  {Saburov}},\ }\href {\doibase 10.1103/PhysRevD.90.012005} {\bibfield
  {journal} {\bibinfo  {journal} {{Phys. Rev. D}}\ }\textbf {\bibinfo {volume}
  {90}},\ \bibinfo {pages} {012005} (\bibinfo {year} {2014})}\BibitemShut
  {NoStop}%
\bibitem [{\citenamefont {Heck}\ \emph {et~al.}(1988)\citenamefont {Heck},
  \citenamefont {Knapp}, \citenamefont {Capdevielle}, \citenamefont {Schatz},\
  and\ \citenamefont {Thouw}}]{CORSIKA}%
  \BibitemOpen
  \bibfield  {author} {\bibinfo {author} {\bibfnamefont {D.}~\bibnamefont
  {Heck}}, \bibinfo {author} {\bibfnamefont {J.}~\bibnamefont {Knapp}},
  \bibinfo {author} {\bibfnamefont {J.~N.}\ \bibnamefont {Capdevielle}},
  \bibinfo {author} {\bibfnamefont {G.}~\bibnamefont {Schatz}}, \ and\ \bibinfo
  {author} {\bibfnamefont {T.}~\bibnamefont {Thouw}},\ }\href
  {http://bibliothek.fzk.de/zb/berichte/FZKA6019.pdf} {\enquote {\bibinfo
  {title} {{CORSIKA: A Monte Carlo Code to Simulate Extensive Air Showers}},}\
  }\bibinfo {type} {FZKA}\ \bibinfo {number} {6019}\ (\bibinfo  {institution}
  {{Forschungszentrum Karlsruhe}},\ \bibinfo {year} {1988})\BibitemShut
  {NoStop}%
\bibitem [{\citenamefont {Kalmykov}, \citenamefont {Ostapchenko},\ and\
  \citenamefont {Pavlov}(1997)}]{QGSJet01}%
  \BibitemOpen
  \bibfield  {author} {\bibinfo {author} {\bibfnamefont {N.~N.}\ \bibnamefont
  {Kalmykov}}, \bibinfo {author} {\bibfnamefont {S.~S.}\ \bibnamefont
  {Ostapchenko}}, \ and\ \bibinfo {author} {\bibfnamefont {A.~I.}\ \bibnamefont
  {Pavlov}},\ }\href {\doibase 10.1016/S0920-5632(96)00846-8} {\bibfield
  {journal} {\bibinfo  {journal} {Nucl. Phys. B - Proc. Suppl.}\ }\textbf
  {\bibinfo {volume} {52}},\ \bibinfo {pages} {17} (\bibinfo {year}
  {1997})}\BibitemShut {NoStop}%
\bibitem [{\citenamefont {Ostapchenko}(2011)}]{QGSJetII}%
  \BibitemOpen
  \bibfield  {author} {\bibinfo {author} {\bibfnamefont {S.}~\bibnamefont
  {Ostapchenko}},\ }\href {\doibase 10.1103/PhysRevD.83.014018} {\bibfield
  {journal} {\bibinfo  {journal} {Phys. Rev. D}\ }\textbf {\bibinfo {volume}
  {83}},\ \bibinfo {pages} {014018} (\bibinfo {year} {2011})}\BibitemShut
  {NoStop}%
\bibitem [{\citenamefont {Ahn}\ \emph {et~al.}(2009)\citenamefont {Ahn},
  \citenamefont {Engel}, \citenamefont {Gaisser}, \citenamefont {Lipari},\ and\
  \citenamefont {Stanev}}]{SIBYLL}%
  \BibitemOpen
  \bibfield  {author} {\bibinfo {author} {\bibfnamefont {E.-J.}\ \bibnamefont
  {Ahn}}, \bibinfo {author} {\bibfnamefont {R.}~\bibnamefont {Engel}}, \bibinfo
  {author} {\bibfnamefont {T.~K.}\ \bibnamefont {Gaisser}}, \bibinfo {author}
  {\bibfnamefont {P.}~\bibnamefont {Lipari}}, \ and\ \bibinfo {author}
  {\bibfnamefont {T.}~\bibnamefont {Stanev}},\ }\href {\doibase
  10.1103/PhysRevD.80.094003} {\bibfield  {journal} {\bibinfo  {journal} {Phys.
  Rev. D}\ }\textbf {\bibinfo {volume} {80}},\ \bibinfo {pages} {094003}
  (\bibinfo {year} {2009})}\BibitemShut {NoStop}%
\bibitem [{\citenamefont {Pierog}\ \emph {et~al.}(2013)\citenamefont {Pierog},
  \citenamefont {Karpenko}, \citenamefont {Katzy}, \citenamefont {Yatsenko},\
  and\ \citenamefont {Werner}}]{EPOS}%
  \BibitemOpen
  \bibfield  {author} {\bibinfo {author} {\bibfnamefont {T.}~\bibnamefont
  {Pierog}}, \bibinfo {author} {\bibfnamefont {I.}~\bibnamefont {Karpenko}},
  \bibinfo {author} {\bibfnamefont {J.~M.}\ \bibnamefont {Katzy}}, \bibinfo
  {author} {\bibfnamefont {E.}~\bibnamefont {Yatsenko}}, \ and\ \bibinfo
  {author} {\bibfnamefont {K.}~\bibnamefont {Werner}},\ }\href@noop {} {\
  (\bibinfo {year} {2013})},\ \Eprint {http://arxiv.org/abs/1306.0121v3}
  {arXiv:1306.0121v3 [hep-ph]} \BibitemShut {NoStop}%
\bibitem [{\citenamefont {Battistoni}\ \emph {et~al.}(2007)\citenamefont
  {Battistoni}, \citenamefont {Muraro}, \citenamefont {Sala}, \citenamefont
  {Cerutti}, \citenamefont {Ferrari} \emph {et~al.}}]{FLUKA:Proc.AIP(2007)}%
  \BibitemOpen
  \bibfield  {author} {\bibinfo {author} {\bibfnamefont {G.}~\bibnamefont
  {Battistoni}}, \bibinfo {author} {\bibfnamefont {S.}~\bibnamefont {Muraro}},
  \bibinfo {author} {\bibfnamefont {P.~R.}\ \bibnamefont {Sala}}, \bibinfo
  {author} {\bibfnamefont {F.}~\bibnamefont {Cerutti}}, \bibinfo {author}
  {\bibfnamefont {A.}~\bibnamefont {Ferrari}},  \emph {et~al.},\ }in\
  \href@noop {} {\emph {\bibinfo {booktitle} {{Proceedings of the Hadronic
  Shower Simulation Workshop 2006}}}},\ Vol.\ \bibinfo {volume} {896},\
  \bibinfo {editor} {edited by\ \bibinfo {editor} {\bibfnamefont
  {M.}~\bibnamefont {Albrow}}\ and\ \bibinfo {editor} {\bibfnamefont
  {R.}~\bibnamefont {Raja}}},\ \bibinfo {organization} {{Fermilab}}\ (\bibinfo
  {publisher} {{AIP Conference Proceeding, New York}},\ \bibinfo {year}
  {2007})\ pp.\ \bibinfo {pages} {31--49}\BibitemShut {NoStop}%
\bibitem [{\citenamefont {Glushkov}, \citenamefont {Diminshtein},\ and\
  \citenamefont {Efimov}(1976)}]{Yakutsk:Ya.Fil.SB.Acad.Sci.USSR(1976)}%
  \BibitemOpen
  \bibfield  {author} {\bibinfo {author} {\bibfnamefont {A.~V.}\ \bibnamefont
  {Glushkov}}, \bibinfo {author} {\bibfnamefont {O.~S.}\ \bibnamefont
  {Diminshtein}}, \ and\ \bibinfo {author} {\bibfnamefont {N.~N.}\ \bibnamefont
  {Efimov}},\ }in\ \href@noop {} {\emph {\bibinfo {booktitle} {{Collection of
  theses}}}}\ (\bibinfo  {publisher} {{Yakutia branch of the Acad. Sci. USSR,
  Yakutsk}},\ \bibinfo {year} {1976})\ p.~\bibinfo {pages} {45},\ \bibinfo
  {note} {{(in Russian)}}\BibitemShut {NoStop}%
\bibitem [{\citenamefont {Ulrich}\ \emph {et~al.}(2001)\citenamefont {Ulrich},
  \citenamefont {Antoni}, \citenamefont {Apel}, \citenamefont {Badea},
  \citenamefont {Bekk} \emph {et~al.}}]{KASCADE:Proc27thICRC(2001)}%
  \BibitemOpen
  \bibfield  {author} {\bibinfo {author} {\bibfnamefont {H.}~\bibnamefont
  {Ulrich}}, \bibinfo {author} {\bibfnamefont {T.}~\bibnamefont {Antoni}},
  \bibinfo {author} {\bibfnamefont {W.~D.}\ \bibnamefont {Apel}}, \bibinfo
  {author} {\bibfnamefont {F.}~\bibnamefont {Badea}}, \bibinfo {author}
  {\bibfnamefont {K.}~\bibnamefont {Bekk}},  \emph {et~al.} (\bibinfo
  {collaboration} {{KASCADE}}),\ }in\ \href@noop {} {\emph {\bibinfo
  {booktitle} {{Proc. of 27th ICRC, Hamburg}}}}\ (\bibinfo  {publisher}
  {{Copernicus, Hamburg, Germany}},\ \bibinfo {year} {2001})\ pp.\ \bibinfo
  {pages} {97--100}\BibitemShut {NoStop}%
\bibitem [{\citenamefont {Budnev}\ \emph {et~al.}(2013)\citenamefont {Budnev},
  \citenamefont {Chernov}, \citenamefont {Gress}, \citenamefont {Korosteleva},
  \citenamefont {Kuzmichev} \emph {et~al.}}]{Tunka:Astropart.Phys.(2013)}%
  \BibitemOpen
  \bibfield  {author} {\bibinfo {author} {\bibfnamefont {N.}~\bibnamefont
  {Budnev}}, \bibinfo {author} {\bibfnamefont {D.}~\bibnamefont {Chernov}},
  \bibinfo {author} {\bibfnamefont {O.}~\bibnamefont {Gress}}, \bibinfo
  {author} {\bibfnamefont {E.}~\bibnamefont {Korosteleva}}, \bibinfo {author}
  {\bibfnamefont {L.}~\bibnamefont {Kuzmichev}},  \emph {et~al.},\ }\href
  {\doibase 10.1016/j.astropartphys.2013.09.006} {\bibfield  {journal}
  {\bibinfo  {journal} {Astropart. Phys.}\ }\textbf {\bibinfo {volume}
  {50-52}},\ \bibinfo {pages} {18} (\bibinfo {year} {2013})}\BibitemShut
  {NoStop}%
\bibitem [{\citenamefont {Andringa}(2010)}]{AUGER(2010)}%
  \BibitemOpen
  \bibfield  {author} {\bibinfo {author} {\bibfnamefont {S.}~\bibnamefont
  {Andringa}} (\bibinfo {collaboration} {{The Pierre Auger Collaboration}}),\
  }\href@noop {} {\  (\bibinfo {year} {2010})},\ \Eprint
  {http://arxiv.org/abs/1005.3795v1} {{arXiv}:1005.3795v1 [hep-ex]}
  \BibitemShut {NoStop}%
\bibitem [{\citenamefont {Abbasi}\ \emph {et~al.}(2010)\citenamefont {Abbasi},
  \citenamefont {Abu-Zayyad}, \citenamefont {Al-Seady}, \citenamefont {Allen},
  \citenamefont {Amman} \emph {et~al.}}]{HiRes:Phys.Rev.Lett.(2010)}%
  \BibitemOpen
  \bibfield  {author} {\bibinfo {author} {\bibfnamefont {R.~U.}\ \bibnamefont
  {Abbasi}}, \bibinfo {author} {\bibfnamefont {T.}~\bibnamefont {Abu-Zayyad}},
  \bibinfo {author} {\bibfnamefont {M.}~\bibnamefont {Al-Seady}}, \bibinfo
  {author} {\bibfnamefont {M.}~\bibnamefont {Allen}}, \bibinfo {author}
  {\bibfnamefont {J.~F.}\ \bibnamefont {Amman}},  \emph {et~al.} (\bibinfo
  {collaboration} {{The High Resolution Fly's Eye Collaboration}}),\ }\href
  {\doibase 10.1103/PhysRevLett.104.161101} {\bibfield  {journal} {\bibinfo
  {journal} {Phys. Rev. Lett.}\ }\textbf {\bibinfo {volume} {104}},\ \bibinfo
  {pages} {161101} (\bibinfo {year} {2010})}\BibitemShut {NoStop}%
\bibitem [{\citenamefont {Tsunesada}(2011)}]{TA(2011)}%
  \BibitemOpen
  \bibfield  {author} {\bibinfo {author} {\bibfnamefont {Y.}~\bibnamefont
  {Tsunesada}} (\bibinfo {collaboration} {{Telescope Array Collaboration}}),\
  }\href@noop {} {\  (\bibinfo {year} {2011})},\ \Eprint
  {http://arxiv.org/abs/1111.2507v1} {arXiv:1111.2507v1 [astro-ph.HE]}
  \BibitemShut {NoStop}%
\end{thebibliography}%

\end{document}